\documentclass[%
 reprint,
superscriptaddress,
nofootinbib,
 amsmath,amssymb
 aps,
longbibliography]{revtex4-1}

\usepackage{amsmath, amssymb}	
\usepackage{bbm}
\usepackage{bm}				
\usepackage{braket}				
\usepackage{graphicx} 			
\usepackage[usenames, dvipsnames]{color}		
\usepackage{mathtools}			
\usepackage{suffix}				
\usepackage{pst-math, pst-plot}	
\usepackage{mathtools}
\usepackage{hyperref}			
\usepackage{cleveref}

\newcommand{\ketbra}[1]{\vert#1\rangle\langle#1\vert}
\DeclareMathOperator{\tr}{tr} 								

\newcommand{\nn}{\nonumber}
\DeclareMathOperator{\Tr}{tr}
\newcommand{\eye}{\mathbb{I}}

\newcommand{\mcF}{\mathcal{F}}

\newcommand{\ve}{\bm{v}}
\newcommand{\Av}{A_{\bm{v}}}
\newcommand{\fd}[1]{\mathbb{#1}}
\newcommand{\Pbv}{\Pi_{a,u}}
\newcommand{\Pbvt}{\Pi^T_{a,u}}
\newcommand{\Pbvp}{\Pi'_{b,v_b}}
\newcommand{\Hd}{\mathcal{H}}
\newcommand{\Hdp}{\mathcal{H}'}
\newcommand{\Hddp}{\mathcal{H}\otimes\mathcal{H}'}
\newcommand{\Den}{\mathcal{D}}
\newcommand{\Denp}{\mathcal{D}'}

\newcommand{\Mo}{M_{\sigma,t}}
\newcommand{\Co}{c_{\sigma,t}}

\newcommand{\Fd}{\fd{Z}_d}

\newcommand{\Fdpdp}{\fd{Z}_{d'}^{d'+1}}
\newcommand{\Free}{\mathcal{F}}
\newcommand{\Freep}{\mathcal{F}'}

\newcommand{\vep}{\bm{u}}
\newcommand{\Avp}{A_{\vep}}
\newcommand{\bb}{b}
\newcommand{\vv}{v}
\newcommand{\Pibv}{\Pi_{\bb,\vv}}
\newcommand{\Zd}{\fd{Z}_d}

\DeclareMathOperator{\conv}{conv}
\newcommand{\qed}{\hfill\square}
\newcommand{\phm}{\phantom{-}}

\renewcommand{\ge}{\geqslant}
\renewcommand{\le}{\leqslant}
\renewcommand{\geq}{\geqslant}
\renewcommand{\leq}{\leqslant}

\newcommand{\bea}{\begin{eqnarray}}
\newcommand{\eea}{\end{eqnarray}}
\newcommand{\be}{\begin{equation}}
\newcommand{\ee}{\end{equation}}
\newcommand{\ba}{\begin{equation}\begin{aligned}}
\newcommand{\ea}{\end{aligned}\end{equation}}

\usepackage{soul}

\begin{document}

\preprint{APS/123-QED}

\title{Quantification and manipulation of magic states}

\author{Mehdi Ahmadi}
\email[]{mehdi.ahmadi@ucalgary.ca}
\affiliation{Department of Mathematics and Statistics, University of Calgary, Calgary, Alberta T2N 1N4, Canada} 
\affiliation{Institute for Quantum Science and Technology, University of Calgary, Alberta T2N 1N4, Canada}

\author{Hoan Bui Dang}
\email[]{hoan.dang@ucalgary.ca}
\affiliation{Department of Mathematics and Statistics, University of Calgary, Calgary, Alberta T2N 1N4, Canada} 
\affiliation{Institute for Quantum Science and Technology, University of Calgary, Alberta T2N 1N4, Canada}

\author{Gilad Gour}
\email[]{gour@ucalgary.ca}
\affiliation{Department of Mathematics and Statistics, University of Calgary, Calgary, Alberta T2N 1N4, Canada} 
\affiliation{Institute for Quantum Science and Technology, University of Calgary, Alberta T2N 1N4, Canada}

\author{Barry C. Sanders}
\email[]{sandersb@ucalgary.ca}
\affiliation{Institute for Quantum Science and Technology, University of Calgary, Alberta T2N 1N4, Canada}
\affiliation{Program in Quantum Information Science, Canadian Institute for Advanced Research,Toronto, Ontario M5G 1Z8, Canada}
\affiliation{Hefei National Laboratory for Physical Sciences at Microscale,University of Science and Technology of China, Hefei, Anhui 230026, China}
\affiliation{Shanghai Branch, CAS Center for Excellence and Synergetic Innovation Center in Quantum Information and Quantum Physics, University of Science and Technology of China, Shanghai 201315, China}

\date{\today}	

\begin{abstract}
Magic states can be used as a resource to circumvent the restrictions due to stabilizer-preserving operations, and magic-state conversion has not been studied in the single-copy regime thus far. Here we solve the question of whether a stabilizer-preserving quantum operation exists that can convert between two given magic states in the single-shot regime. We first phrase this question as a feasibility problem for a semi-definite program (SDP), which provides a procedure for constructing a stabilizer-preserving quantum operation (free channel) if it exists. Then we employ a variant of the Farkas Lemma to derive necessary and sufficient conditions for existence, and this method is used to construct a complete set of magic monotones.\end{abstract}

\maketitle


\section{Introduction}
\label{Introduction}

Magic states have interesting applications in disparate areas of quantum physics from foundations of quantum mechanics to quantum computation. According to the Gottesman-Knill's theorem \cite{GottesmanPhD, Bartlett2002,Mari2012}, if we restrict the allowed set of states and operations to stabilizer states and operations, then the dynamics and measurements of quantum states can be simulated efficiently on a classical computer. However, it is well-known that universal quantum computing can be achieved by addition of magic states~\cite{Bravyi2005,Howard2016}. In \textsl{circuit synthesis} \cite{Howard2016}, the set of Clifford unitaries is supplemented with the $T$-gate in order to achieve universal quantum computation. There, it is favorable to reduce the number of $T$-gates as much as possible, since to implement a $T$-gate, magic states need to be consumed as a resource. Moreover, recently an interesting connection between \textsl{contextuality} and the resource theory of magic has been pointed out~\cite{Spekkens2008,Delfosse2015,Raussendorf2016,Delfosse2016}. In particular, it was established that quantum contextuality is a resource for quantum speed-up within one of the most successful models for fault-tolerant quantum computation~\cite{Howard2014}, namely the magic state distillation model.\\

Recently, the resource theory of magic has attracted much attention~\cite{Veitch2014, Howard2014, Delfosse2015, Delfosse2016, Raussendorf2016}. In this framework, free operations are the set of allowed operations, i.e., stabilizer operations. Resource states, namely the magic states, are required in order to achieve some desired task. In a realistic setting wherein the resources are finite, one is ideally interested in answering the single-shot question: Given two resource states $\rho$ and $\rho'$, is there a free operation that will convert $\rho$ to $\rho'$? In a recent attempt to answer this question, the necessary and sufficient conditions for the possibility of converting a resource state into another resource state has been obtained for a large class of quantum resource theories coined as \textsl{affine resource theories}~\cite{GiladART}. In these resource theories, an affine combination of free states is considered to be a free state itself. Resource theories of coherence~\cite{Baumgratz2014,Chitambar2016a,Chitambar2016b,Marvian2016,Winter2016,Napoli2016}, asymmetry~\cite{Gour2008,Gour2009,Marvian2013, Ahmadi2013,Marvian2014a,Marvian2014b}, athermality~\cite{Brandao2013,Brandao2015,Horodecki2013,Faist2015,Lostalgio2015a,Lostalgio2015b,Gour2015a} are examples of affine resource theories, whereas the resource theories of entanglement and \textsl{magic states}~\cite{Veitch2014,Howard2014,Howard2016} are not affine~\cite{GiladART}.\\

In this paper, we study the single-shot conversion of magic states using free operations. We find the set of free operations in \cite{Veitch2014} too restrictive and we extend it to include all the completely positive trace preserving (CPTP) maps that convert stabilizer states into stabilizer states, i.e., \textsl{stabilizer-preserving operations} (SPOs). SPOs form the largest possible set of physical operations that can be considered free. We have numerical evidence that this set is strictly larger than the set of stabilizer operations as defined in \cite{Veitch2014}. Furthermore, we construct a complete set of magic monotones based on the conditional min-entropy~\cite{Konig2009}. Our set of magic monotones is complete in the sense that a magic state can be converted into another magic state by an SPO if and only if the value of all the monotones does not increase in the process.\\

This paper is structured as follows: In section~\ref{FS}, we define the set of free states, i.e., stabilizer states. In section~\ref{FO}, we characterize the set of SPOs. In section~\ref{monotones}, we construct a family of magic monotones which quantify the usefulness of magic states. In section~\ref{CNSC}, we formulate the necessary and sufficient conditions for single-shot conversion of a magic state to another using a free operation as an SDP.  In section~\ref{complete}, we prove that our set of magic monotones is complete.\\

Throughout this paper, $d$ denotes the dimension of the Hilbert space. $\Hd$ is the set of $d\times d$ Hermitian matrices and $\Den$ is the set of $d\times d$ density matrices. When a quantum channel is considered, we use no-prime notation for the input of the channel and prime notations for the output (for example, $\Free$ denotes the set of free input states in dimension $d$ while $\Freep$ denotes the set of free output states in dimension $d'$). Greek letters are used for density matrices. Capital letters such as $X,Y$ and $Z$ are used for generic Hermitian matrices, with an added tilde if they are traceless.
\section{Free states}\label{FS}
The free states in our resource theory are defined to be stabilizer states. These consist of all pure stabilizer states, which are eigenstates of the generalized Pauli operators, and their convex mixtures. We use the term magic (or non-stabilizer) states to refer to states that are not stabilizer states. We only consider cases where the Hilbert space dimension $d$ is a prime number. We devote a separate analysis for $d=2$ to provide intuition via the visualization of the free states in the Bloch sphere.
\subsection{Qubit case}
\begin{figure}[t]\centering
\includegraphics[scale = 0.28]{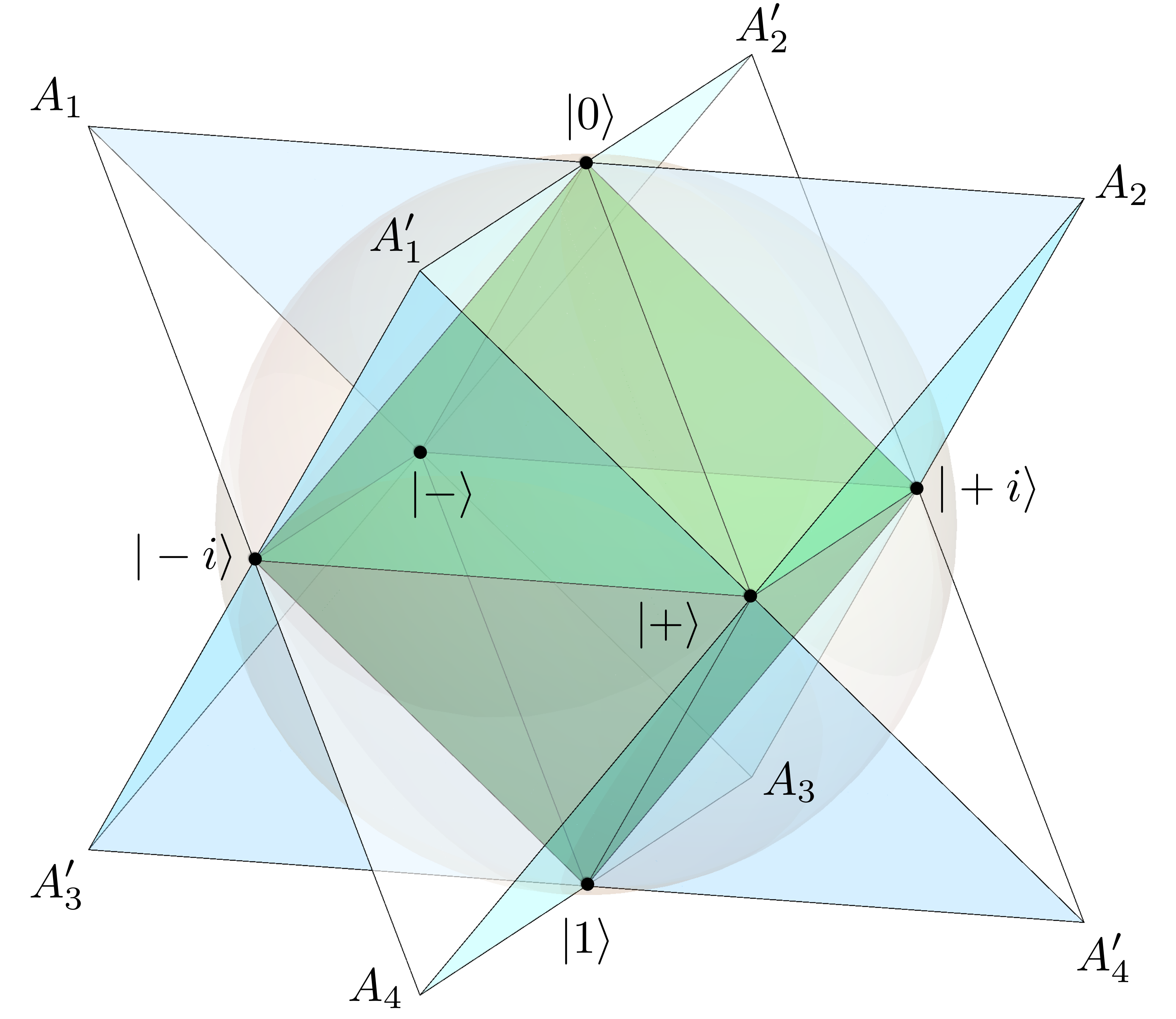}
\vskip 2mm
\parbox{\linewidth}{\caption[]{The octahedron of stabilizer states for a qubit as an intersection of two regular tetrahedra. The 6 pure stabilizer states $\{\ket{0},\ket{1},\ket{+},\ket{-},\ket{i}, \ket{-i}\}$ are vertices of the octahedron, which lie on the Bloch sphere. The phase-point operators $\{A_i\}$ and $\{A_i'\}$ are the vertices of the two tetrahedra.}
\label{fig1}}
\end{figure}

In the Hilbert space of dimension $d=2$, there are 6 pure stabilizer states, namely the set of eigenstates of Pauli operators $\{\ket{0},\ket{1},\ket{+},\ket{-},\ket{i}, \ket{-i}\}$. In the Bloch representation, these states correspond to 3 pairs of antipodal points along the 3 principle axes. The full set of free qubit states $\mcF$ is the convex hull of these extreme points, which is an octahedron embedded in the Bloch sphere. This octahedron consists of 4 pairs of parallel facets, so a point lies inside the octahedron if and only if it is confined to the space between any pair of parallel planes containing the facets. These planes are described by the equations $\pm x \pm y \pm z = 1$. Therefore, a state with the Bloch vector $(x,y,z)$ is a free state if and only if the following inequalities hold:
\begin{equation}\label{ineq-free1}
\begin{split}
-1 &\le \phantom{-}x + y + z \phantom{-}\le 1 \\
-1 &\le -x + y + z \phantom{-}\le 1 \\
-1 &\le \phantom{-}x - y + z\phantom{-} \le 1 \\
-1 &\le \phantom{-}x + y - z \phantom{-}\le 1. \\
\end{split}
\end{equation}
The inequalities above characterize free qubit states in terms of their Bloch coordinates. However, as it will be more natural for the later consideration of quantum channels, we would like to have an alternative characterization in the space of Hermitian operators. This can be done by noticing that the facets of the octahedron can be divided into 2 groups, each containing 4 facets that extend to a regular tetrahedron. The octahedron can then be described as the intersection of these 2 regular tetrahedra as shown in Figure \ref{fig1}.\\

The 8 vertices of the 2 tetrahedra have Bloch coordinates $(\pm1,\pm1,\pm1)$, where the first tetrahedron's vertices consist of coordinates with an odd number of $1$'s such as $(1,1,1)$ and $(-1,-1,1)$, and the second tetrahedron's vertices have an even number of $1$'s in their coordinates. Let $A_i$ (for $i=1,\dots,4$) denote the unit-trace Hermitian operators corresponding to the vertices of the first tetrahedron (similarly $A'_i$ for the second one). It is straightforward to check that
\begin{equation}\label{eq-TrA}
\Tr{A_i A_j} = \begin{cases}
2 & \text{if }i=j\\
0 & \text{if }i\ne j.\\
\end{cases}
\end{equation}
The same properties hold for $A'_i$. Therefore, $\{A_i\}$ and $\{A'_i\}$ form two orthogonal (but not normalized) bases for the space of qubit Hermitian operators. Any unit-trace Hermitian operator $H$ can therefore be written as
\begin{equation}
H = \sum_{i=1}^4 \alpha_i A_i = \sum_{i=1}^4 \alpha_i' A_i'
\end{equation}
where the real coefficients $\alpha_i$ and $\alpha_i'$ are determined by
\begin{equation}
\alpha_i = \frac{1}{2} \Tr{(H A_i)}   \hskip 10mm \alpha_i' = \frac{1}{2} \Tr{(H A_i')}
\end{equation}
and they satisfy $\sum_i \alpha_i =\sum_i \alpha_i' = 1$ because $H$ has unit trace. In Bloch space, $H$ lies inside the tetrahedron $\{A_i\}$ if and only if it can be written as a convex combination of $A_i$, meaning that the coefficients $\alpha_i$ are all non-negative. A similar statement holds for the tetrahedron $\{A_i'\}$. This results in the following characterization of free states: a state $\rho$ is free if and only if
\begin{equation}\label{ineq-free2}
\begin{split}
\Tr (\rho A_i) &\ge 0,   \\\
\Tr (\rho A_i') &\ge 0  
\end{split}
\end{equation}
for all $i = 1,\dots,4$. \\

Note that the number of inequalities in (\ref{ineq-free2}) is the same as that in (\ref{ineq-free1}), but here we can directly characterize the set of free states $\mcF$ using the Hilbert-Schmidt inner product in the space of operators. It is also worth noting that $\{\alpha_i\}$ and $\{\alpha_i\}$ are sets of values for the Wigner functions associated with the phase-point operators $\{A_i\}$ and $\{A_i'\}$.\\

\subsection{Qudit case}
Next, we define and characterize free states for the case of a single qudit of dimension $d$, where $d$ is an odd prime number. Following the same procedure for the qubit case, we will start with the definition of pure stabilizer states, and then take the convex hull to obtain the set of all free states $\Free$. Similar to the qubit case, we can characterize $\Free$ by the simultaneous positivity of a finite number of Wigner functions, as shown below.\\

Pure stabilizer states can be obtained by applying Clifford unitaries to the state $\ket{0}$ \cite{Veitch2014}. In prime dimensions, they  form a complete set of Mutually Unbiased Bases (MUBs) \cite{Hoanthesis}. In other words, they can be grouped into $d-1$ orthonormal bases so that states from different bases all have the same overlap. If we denote the density operators of these states by $\Pibv$, where $\bb \in \{1,\dots,d+1\}$ labels the basis and $\vv \in \{0,\dots,d-1\}$ labels the basis vector, then they form a complete set of MUBs if and only if they are rank-1 projectors that satisfy
\be \label{MUBproperties}
\tr (\Pi_{\bb,\vv} \Pi_{\bb',\vv'}) = 
\begin{cases}
1 &\text{if } \bb = \bb', \vv = \vv' \\
0  &\text{if } \bb = \bb', \vv \ne \vv' \\
1/d  &\text{if } \bb \ne \bb'.
\end{cases}
\ee
The set of all (pure and mixed) stabilizer states is then defined as the convex hull of the pure stabilizer states, which is denoted by $\Free := \conv\{\Pibv\}$.\\

The standard discrete Wigner function of a quantum state $\rho \in \Den$ is a quasi-probability distribution over the discrete phase space $\Zd \times \Zd$ given by
\be
W_{p,q}(\rho) := \tr (\rho A_{p,q}),
\ee
where $(p,q)\in \Zd \times \Zd$ and $A_{p,q}$ are $d^2$ Hermitian operators that satisfy
\be
\begin{split}
\tr A_{p,q} &=1\\
\tr (A_{p,q} A_{p',q'}) &= d \delta_{pp'}\delta_{qq'}.
\end{split}
\ee
$A_{p,q}$ are usually referred to as phase-point operators. For odd prime dimensions $d$, they are defined as
\be
A_{0,0} := \frac{1}{d}\sum_{p,q} D_{p,q}, \hskip 5mm A_{p,q} := D_{p,q} A_{0,0} D_{p,q}^{\dagger}
\ee
where 
\be
D_{p,q} := \omega^{-pq/2}P^p S^q, \hskip 5mm \omega := e^{2\pi i/d}
\ee
and the shift operator $S$ and phase operator $P$ are defined by their actions on the standard basis as
\be 
S\ket{k} = \ket{k+1}, \hskip 5mm P\ket{k} = \omega^k\ket{k}, \hskip 5mm k \in \Zd.
\ee

The discrete Hudson's theorem proves that a pure state is a stabilizer state if and only if its standard Wigner function is non-negative for all $(p,q) \in \Zd \times \Zd$ \cite{Gross}. This is not true for mixed states: there are mixed states with non-negative standard Wigner function that are not stabilizer states (i.e. not in $\Free$)~\cite{Gross}. However, if we consider a larger family of $d^{d-1}$ Wigner functions, it was conjectured \cite{Galvao2005} and later proved \cite{Cormick2006} that $\Free$ is characterized by the simultaneous non-negativity of these Wigner functions. For the purpose of characterizing $\Free$, we do not need to go into the details of how to define them, and will instead only specify their associated $d^{d+1}$ generalized phase-point operators.\\

Let $\ve = (\vv_1, \vv_2,\dots,\vv_{d+1})$ be a vector with $d+1$ components $\vv_i \in \Zd$. For each $\ve$ we define a generalized phase-point operator $\Av$ as
\be
\Av := \sum_{b=1}^{d+1} \Pi_{\bb,\vv_{\bb}} - \eye.
\ee
There are $d^{d+1}$ such generalized phase-point operators. It is straightforward to check from their definition that $\tr(\Av) = 1$. Using the properties of MUB projectors in (\ref{MUBproperties}), one can show that $\tr(\Av \Avp)=d$ if and only if $\ve$ and $\vep$ agree at exactly one component. One can group these operators into $d^{d-1}$ groups of size $d^2$ in such a way that for any $\Av$ and $\Avp$ in the same group, $\ve$ and $\vep$ agree at exactly one component, thus forming $d^{d-1}$ Wigner functions (this is a non-trivial combinatoric problem, see section 4 in \cite{Bengtsson2005}). Here, we focus on the fact that these operators are stabilizer witnesses and we can use them to characterize $\Free$: a density operator $\rho$ is a free state if and only if
\be \label{ineq-freestate}
\tr(\rho \Av) \ge 0 \hskip 3mm \forall \ve \in \Zd^{d+1}.
\ee
A geometric interpretation of the inequalities in \eqref{ineq-freestate} is that $\Free$ is a polytope with $d^{d+1}$ facets in the space of Hermitian operators.

\section{Free operations}\label{FO}

In magic-state distillation protocols, one asks whether there exists a procedure, which can transform any mixed magic state to a more resourceful (more magic) state while solely employing free operations.  In \cite{Veitch2014}, stabilizer protocols and magic monotones such as sum negativity, mana and relative entropy of magic were defined to answer this question quantitatively. In this section, for comparison purposes, we first recall the definition of stabilizer protocols and then present our definition of free operations (SPOs).\\

A stabilizer protocol may consist of the following types of quantum operations: Clifford unitaries, composition with stabilizer states, partial trace, and measurements in the computational basis. Moreover, these quantum operations can be conditioned on measurement results and classical randomness. Using the Stinespring dilation theorem, any quantum protocol composed of these quantum operations can be written as
\begin{align}\label{dilation}
	{\cal E}(\rho)=\Tr_E\left[U\left(\rho\otimes\rho_{E}\right)U^{\dag}\right],
\end{align}
where $U$ is a Clifford unitary\footnote{The Clifford group is defined as the normalizer of the generalized Pauli group, i.e., the collection of unitary operators $U_C$ that map
the generalized Pauli group to itself under conjugation.} and the ancilla $\rho_{E}$ is a free state, i.e., a pure or mixed stabilizer state. Note that for a given dimension of the ancillary system there are finitely many Clifford unitaries which means that the number of stabilizer protocols \eqref{dilation} is also finite.\\
  
In our definition of SPOs as free operations, we do not put any operational restriction on $U$ and $\rho_E$ but instead only require that free operations ${\cal O}_f$ do not generate resources. Specifically, we define SPOs as linear maps that satisfy the following conditions:
\begin{enumerate}
\item They are completely positive (CP),
\item They are trace preserving (TP),
\item They transform stabilizer states to stabilizer states; i.e., if $\rho \in \Free$, then ${\cal E} (\rho) \in \Freep$.
\end{enumerate} 
We would like to point out that in the resource theory framework the set of SPOs as defined above is the maximal set of resource non-generating operations. Therefore, it includes the stabilizer protocols. In fact, we have numerically checked that the set of SPOs strictly contains the set of stabilizer protocols, as described in the appendix.\\
      
We write the above conditions (1-3) in the Choi representation which later enables us to phrase our problem as an SDP. The Choi matrix $J$ corresponding to the quantum channel ${\cal E}: \Hd\to\Hdp$ is given as $J={\cal E}\otimes \mathbb{I}(\ket{\phi^+}\bra{\phi^+})$ where $\ket{\phi^+}=\sum_{i=1}^{d}\ket{ii}$ which is the state vector of the maximally entangled state up to a normalization, and $J$ is a positive semi-definite matrix in $\Hddp$. Note that the TP condition can be phrased in terms of the Choi matrix as
$\tr_B(J)=\eye$ which is equivalent to $\tr\left(J(\eye'\otimes X)\right)=\Tr(X)$ for all $X\in \Hd$. Finally, the Choi matrix of a free operation ${\cal E}\in {\cal O}_f$ must satisfy the conditions

\begin{align}
&J\geq0,\label{CP1}\\
&\tr\left(J(\eye'\otimes X)\right)=\Tr(X)\;\;\; \forall X\in \Hd,\label{TP1}\\
&\tr\left(J \left(\Av \otimes \Pbvt\right)\right)\geq0 \;\;\;\forall \ve,a,u,\label{F1}\end{align}
where $\ve\in\Fdpdp$, $a=1,2,\ldots,(d+1)$ and $u\in\Fd$.\\

Note that to write the third condition, given in Eq.~\eqref{F1}, we have used the fact that, if a CPTP map takes the extreme points of the stabilizer polytope to a free state, it will also take any convex combination of them to a free state. Hence, it is sufficient to demand that the quantum channel takes all the elements of the set of MUB projectors $\{\Pbv\}$ to a free state.\\

\section{Quantifying magic}\label{monotones}
Monotones are used to quantify the resourcefulness of resource states. By definition, they are real-valued functions that are non-increasing under the action of free operations. In other words, for $f$ to be a monotone, it has to satisfy $f(\rho) \ge f({\cal E}(\rho))$ for all states $\rho$ and free operations ${\cal E}$. In this section, we construct a family of magic monotones to quantify magic states and prove that they are indeed non-increasing under SPO.\\

{\bf Definition: } For any non-zero $X \in \mathcal{H}^A \otimes \mathcal{H}^B$, $X \ge 0$, we define $q(X)$ as
\be
q(X) :=\min_{S\ge 0} \{ \tr S | S \in \mathcal{H}^B, \eye^A \otimes S \ge X \}.
\ee
Note that $q(X)$ is positive and closely related to the conditional min-entropy by $q(X) = 2^{-H_{\text{min}}(A|B)_X}$ \cite{Konig2009}.\\

Given a density matrix $\sigma \in \Denp$, a real number $t \ge 0$ and probabilities $p_{\ve,a,u}$, we define the density matrix $\Omega_{\rho}$ as a function of input state $\rho \in \Den$:
\be \label{def-Omega}
\Omega_{\rho}:=N\left( \sigma \otimes \rho^T + t\sum_{\ve,b,a,u}p_{\ve,a,u}\Pbvp  \otimes \Pbvt \right) 
\ee
where $N:=1/(td'+t+1)$ is the normalization factor and the sum is over all $\ve\in\Fdpdp$, $b=1,\dots,d'+1$, $u\in\Fd$ and $a=1,\dots,d+1$.\\

{\bf Definition (monotones):} For each $\sigma \in \Denp$ and a real number $t \ge 0$, we define the function $\Mo:\Den \to \mathbb{R}$ as
\be\label{def-monotone}
\Mo := \min_{p_{\ve,a,u}} q(\Omega_{\rho}) - \Co 
\ee
where $\Co:= \min_{p_{\ve,a,u}} q(\Omega_{\eye/d})$.\\
\begin{figure}[t]\centering
\includegraphics[scale = 0.4]{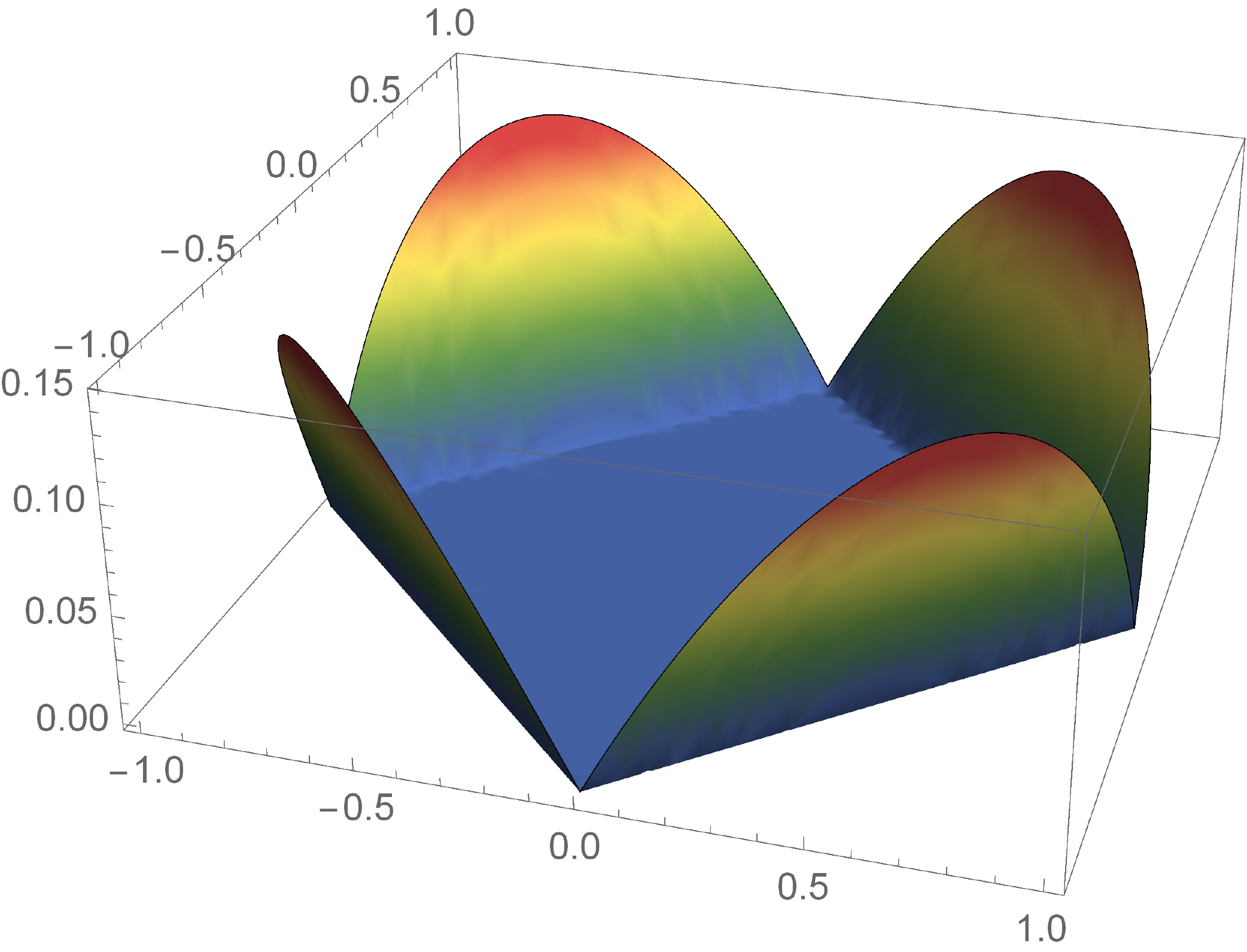}
\vskip 2mm
\parbox{\linewidth}{\caption[]{Plot of $M_{\ketbra{0},1}(\rho)$ evaluated at qubit states $\rho$ on the equatorial disk of the Bloch sphere. The vertical axis represents the value of the monotone. The horizontal axes represent the $x$ and $y$ Bloch coordinates of $\rho$. The flat square in the center corresponds to stabilizer states.} 
\label{fig2}}
\end{figure}

Note that $\Omega_{\rho}$ depends on $\sigma$, $t$ and probabilities $p_{\ve,a,u}$, but $\Mo$ only depends on parameters $\sigma$ and $t$ because of the minimization over $p_{\ve,a,u}$. The constant $\Co$ is used to offset the value of $\Mo$ to zero on stabilizer states.\\

{\bf Theorem 1:} For all $\sigma \in \Denp$ and $t \ge 0$, $\Mo(\rho)$ are magic monotones. That is, for an arbitrary state $\rho$, if $\rho' = \cal{E}(\rho)$ for a stabilizer-preserving quantum channel $\cal{E}$, then
\be \Mo(\rho) \ge \Mo(\rho'). \ee
{\it Proof:} We first recall a fact about the conditional min-entropy that for a channel $\cal{E}$,   $q(X)\geq q(\Lambda(X))$, where $\Lambda := \mathbb{I}\;\otimes\; T \circ {\cal E} \circ T$ and $T$ is the transposition map \cite{GiladART}. Also, note that because $\cal{E}$ is a stabilizer-preserving channel,
\begin{align}
{\Lambda}\left(\Pbvp  \otimes \Pbvt\right)&=\Pbvp  \otimes {\cal E}(\Pbv)^T\nonumber\\
&=\Pbvp \otimes\sum_{a,u}p_{a,u}\Pbvt
\end{align}
for some probabilities $p_{a,u}$. Denoting the set of probabilities that minimize the function $q\left( \Omega_{\rho}\right)$ by $\tilde{p}_{\ve,a,u}$, we then have 
 \begin{align}
 &M_{\sigma,t}(\rho) + \Co=\nn\\
 & q\left(N\left( \sigma \otimes \rho^T + t\sum_{\ve,b,a,u}\tilde{p}_{\ve,a,u}\Pbvp  \otimes \Pbvt \right)\right)\nn \\ \nn &\geq 
 q\left(N\left( \sigma \otimes \rho'^T + t\sum_{\ve,b,a,u}\bar{p}_{\ve,a,u}\Pbvp  \otimes \Pbvt \right)\right)\nn\\ &\geq  M_{\sigma,t}(\rho') + \Co 
 \end{align}
where $N:=1/(td'+t+1)$. The first inequality follows from $q(X)\geq q(\Lambda(X))$, $\bar{p}_{\ve,a,u}$ are probabilities derived from $\tilde{p}_{\ve,a,u}$ and $p_{a,u}$, and the second inequality follows from the definition of $\Mo$.  $\qed $\\

{\it Remark 1.} Our monotones are also monotones under the set of operations considered in~\cite{Veitch2014}, since it is a strict subset of our set of SPO.\\

{\it Remark 2.} It is worth pointing out that the double minimization in the definition of $\Mo$ can be computed in whole as an SDP. As the number of probabilities $p_{\ve,a,u}$ grows as $d(d+1)d'^{d'+1}$, in high dimensions it is not practical to compute $\Mo$. The computation can be easily carried out in low dimensions, as described in the following numerical example.\\

\begin{figure}[t]\centering
\includegraphics[scale = 0.63]{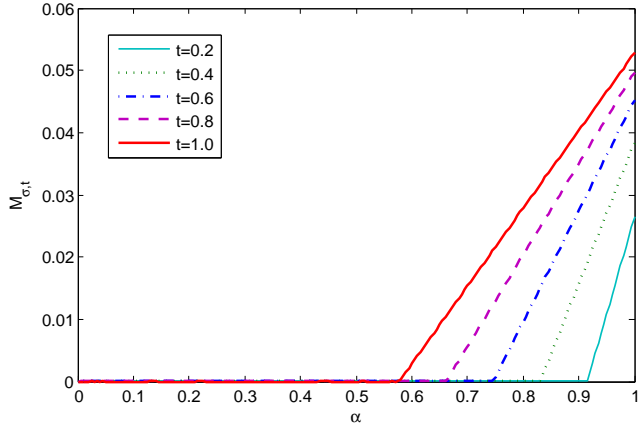}
\vskip 2mm
\parbox{\linewidth}{\caption[]{Plot of $M_{\ketbra{0},t}(\rho)$ for several values of $t$, evaluated at qubit states $\rho = (1-\alpha)\eye/2 + \alpha \ketbra{T}$
 along the line from the completely mixed state to the type-T magic state with Bloch coordinate $(1,1,1)/\sqrt{3}$. The boundary between stabilizer and non-stabilizer states is at $\alpha \approx 0.58$. The monotone can be seen to be faithful when $t=1$.}
\label{fig3}}
\end{figure}

{\bf Example (qubit case):} Here we consider the case $d=d'=2$. Computing $\Mo$ involves a minimization over 48 real parameters and a $2\times 2$ Hermitian matrix. We used CVX in Matlab to evaluate $\Mo(\rho)$ numerically at various input states $\rho$, for several values of parameters $\sigma$ and $t$. With $\sigma$ fixed at $\ketbra{0}$, for $t=0$ we get the trivial zero monotone. For $t=1$ the monotone is faithful, meaning that $\Mo(\rho) = 0$ if and only if $\rho$ is a stabilizer state (see Figure \ref{fig2}). For $0\le t<1$ or $t>1$, the monotone is not faithful: there are magic states $\rho$ at which $\Mo(\rho)=0$ (see Figure \ref{fig3}). For some other choices of $\sigma$, we observed that the monotone became faithful at some particular value of $t$ that depends on $\sigma$.

\section{Manipulating magic states}\label{MMS}
Resource state manipulation, i.e.,  using free operations to convert one resource state to another, is among the most fundamental aspects in studying a quantum resource theory. Here we consider exact manipulation in the single-shot regime, that is, exact conversion of a single copy of one state to that of another by an SPO. We first formulate the conversion question as a semi-definite program and use Farkas' lemma to derive necessary and sufficient conditions for conversion. We then show that the monotones constructed in the previous section can alternatively be used to determine convertibility.
\subsection{An SDP formulation}\label{CNSC}
The characterization of free operations in \eqref{CP1}-\eqref{F1} is readily in the form of SDP constraints. The only missing ingredient is the condition for a quantum channel ${\cal E}$ to convert state $\rho$ to state $\rho'$. We begin this section by expressing the state conversion statement ${\cal E}(\rho)=\rho'$ in the Choi matrix representation as $\tr_{B}\left(J\left(\mathbb{I}\otimes\rho^{T}\right)\right)=\rho'$. After transforming the partial-trace to a full trace we get  
\begin{align}\label{C1}
\tr\left(J\left(Y\otimes\rho^{T}\right)\right)=\tr(\rho' Y) \;\;\;\forall Y\in \Hdp.
\end{align}
Following \cite{GiladART}, we can phrase the existence question of a stabilizer-preserving channel ${\cal E}$ converting density matrices $\rho$ to $\rho'$ as the existence question of a non-zero matrix $J\in \Hddp$ that satisfies the conditions:
\begin{subequations}
\label{SDP}
\begin{align}
&J\geq 0\label{CP3}\\
&\tr\left(J(\eye'\otimes \tilde{X})\right)=0\;\;\; \forall \tilde{X}\in \Hd\label{TP3}\\
&\tr\left(J \left(\Av \otimes \Pbvt\right)\right)\geq0 \;\;\;\forall \ve,a,u, \label{F3}\\
&\tr\left(J\left(\tilde{Y}\otimes\rho^{T}-\frac{\tr(\tilde{Y} \rho')}{d} \,\eye'\otimes\eye\right)\right)=0 \;\;\forall \tilde{Y}\in \Hdp,\label{SC3}
\end{align}
\end{subequations}
for all traceless matrices $\tilde{X}$ and $\tilde{Y}$, $\ve\in\Fdpdp$, $u\in\Fd$ and $a=1,\dots,d+1$. Conditions \eqref{SDP} on the Choi matrix $J$ correspond to the conditions for ${\cal E}$ to be completely-positive, trace-preserving, stabilizer-preserving, and converting $\rho$ to $\rho'$. To turn the existence question into necessary and sufficient conditions that are verifiable, we make use of the following lemma.
\\

{\bf Lemma 1: } Let $\mathcal{V}$ be a subspace of $\Hd$ and let $W_1,\dots,W_m$ be matrices in $\Hd$. The following are equivalent.
\begin{enumerate}
\item There exists a non-zero positive semidefinite matrix $J$ such that
\begin{equation}
\begin{split}
&\tr(J V) = 0 \hskip 6mm \text{ for all } V \in \mathcal{V} \\
&\tr(J W_i) \ge 0 \hskip 5mm \text{ for all } i=1,\dots,m.
\end{split}
\end{equation}
\item For all $V \in \mathcal{V}$ and non-negative numbers $y_1,\dots,y_m$
\begin{equation}
V - (y_1 W_1 + \dots + y_m W_m) \ngtr 0.
\end{equation}
\end{enumerate}

This lemma is a variant of Farkas' Lemma and its proof is based on the separating hyperplane theorem~\cite{Boyd2004}. Applying this lemma to conditions \eqref{CP3}-\eqref{SC3}, with $\mathcal{V}$ being the subspace spanned by $\eye'\otimes \tilde{X}$ and $ \tilde{Y} \otimes \rho^T- \frac{\tr(\tilde{Y}\rho')}{d} \eye' \otimes \eye$, and $W_i$ being $\Av \otimes \Pbvt $, we obtain the following equivalent conditions:
\begin{equation}\label{NSC}
\eye'\otimes \tilde{X} + \tilde{Y} \otimes \rho^T- \frac{\tr(\tilde{Y}\rho')}{d} \eye' \otimes \eye - \sum_{\ve,a,u} y_{\ve,a,u} \Av \otimes \Pbvt \ngtr 0
\end{equation}
for all traceless $\tilde{X} \in \Hd$, traceless $\tilde{Y} \in \Hdp$, and $y_{\ve,a,u} \ge 0$. \\

To summarize, we have cast the state conversion question in the form of an SDP feasibility problem \eqref{CP3}-\eqref{SC3} and alternatively provided a set of necessary and sufficient conditions. The SDP form is more useful for practical purposes. It is also constructive: when the conversion is possible, there are known SDP algorithms that can find an SPO that does the state conversion. The necessary and sufficient conditions in \eqref{NSC} consist of infinitely many conditions. On the other hand, they provide the analytical base for the proof of the completeness of our magic monotones in the next section.\\

{\it Remark 3.} Using CVX in Matlab, we found a numerical example of qutrit-to-qutrit SPO that can increase the Wigner sum negativity\footnote{The sum negativity of a state is the sum of the negative elements of the Wigner
function~\cite{Veitch2014}.} (a monotone for stabilizer protocols studied in \cite{Veitch2014}), thus confirming that the set of SPOs is strictly larger than the previously studied set of stabilizer protocols. See the appendix for this specific example.


\subsection{Completeness of monotones}\label{complete}
An alternative way of checking if a resource can be converted to another by a free operation is to use a complete set of monotones. In this section, we show that our set of magic monotones, $\Mo(\rho)$ defined in \eqref{def-monotone}, is complete in the sense that a resource state can be converted to another using a stabilizer-preserving operation if and only if it is more resourceful with respect to every monotone. We begin with the lemma below, which we then use to prove the completeness of magic monotones as stated in Theorem 2. \\

{\bf Lemma 2:} Given density operators $\rho \in \Den$ and $\rho' \in \Denp$, there exists a stabilizer-preserving quantum channel $\mathcal{E}$ such that $\mathcal{E}(\rho) = \rho'$ if and only if for all density operators $\sigma \in \Denp$, probabilities $p_{\ve,a,u}$ and $t \ge 0$ we have 
\begin{align}
\frac{t+\tr(\sigma\rho')}{td'+t+1}\leq q\left(\Omega_{\rho}\right).
\end{align}
{\it Proof:} The necessary and sufficient conversion conditions were cast in (\ref{NSC}). Notice that these conditions remain unchanged if we divide the left hand side by an arbitrarily large positive number. Therefore, we can assume $Y < \eye'/d'$ such that $Y = \eye'/d' - \sigma$, where $\sigma$ is a density operator. After making this substitution into (\ref{NSC}), diving both sides by $t(d'+1)+1$ and re-arranging it so that all the terms of the form $\eye' \otimes \ldots$ are grouped to the left-hand side of the inequality, we obtain
\begin{equation}
\begin{split}
\eye' \otimes \tau \ngtr \Omega_{\rho}\end{split}
\end{equation}
where $ \Omega_{\rho}$ is defined in \eqref{def-Omega} and 
\begin{align}
&\tau :=\\
&\frac{1}{td'+t+1}\left(X + \frac{\rho^T}{d'} + \frac{\tr(\sigma\rho')}{d}\eye -\frac{\eye}{dd'}+t \sum_{a,u} p_{a,u}\Pbvt\right)	\nn
\end{align}
for all traceless $\tilde{X}\in \Hd$, $\sigma \in \Denp$ , probabilities $p_{a,u}$ and $t \ge 0$. This is equivalent to 
\begin{align}
\tr(\tau)=\frac{t+\tr(\sigma\rho')}{td'+t+1}\leq q(\Omega_{\rho})
\end{align}
for all density operators $\sigma \in \Denp$ and $t\ge 0$. $\qed$\\

{\bf Theorem 2:} Given density operators $\rho \in \Den$ and $\rho' \in \Denp$, if for all density operators $\sigma \in \Denp$ and $t \ge 0$ it holds that $\Mo(\rho) \ge \Mo(\rho')$, then there exists a stabilizer-preserving quantum channel $\mathcal{E}$ such that $\mathcal{E}(\rho) = \rho'$. \\



{\it Proof:}  We begin by noticing that the identity channel is a free channel, and therefore we know that $\rho'$ can be converted to $\rho'$. Then using lemma (2) and the optimal set of probabilities $\tilde{p}_{\ve,a,u}$, for all density operators $\sigma \in \Denp$ and $t \ge 0$ we can write 
\bea 
&&\frac{t+\tr(\sigma\rho')}{t(d'+1)+1}\nn\\
&\leq &\frac{1}{td'+t+1} ~q\left( \sigma \otimes \rho'^T + t\sum_{\ve,b,a,u}\tilde{p}_{\ve,a,u}\Pbvp  \otimes \Pbvt\right)\nn\\
&=&\Mo(\rho') \leq \Mo(\rho)\le q\left(\Omega_{\rho}\right),\nn\eea
and hence there exists a free channel which converts $\rho$ to $\rho'$. $\qed$\\

The following corollary is the result of combining theorems (1) and (2).\\

{\bf Corollary:} Given density operators $\rho \in \Den$ and $\rho' \in \Denp$, the following statements are equivalent:
\begin{enumerate}
\item There exists a stabilizer-preserving quantum channel $\mathcal{E}$ such that $\mathcal{E}(\rho) = \rho'$
\item For all density operators $\sigma \in \Denp$ and $t \ge 0$ we have
$\Mo(\rho) \ge \Mo(\rho')$.
\end{enumerate}
To summarize, we have found a complete set of magic monotones which can be used to determine the single-shot convertibility between two states.

\section{Conclusions}\label{SD}
In this paper, we answered the question: Given two magic states $\rho$ and $\rho'$, is there an SPO that can convert $\rho$ to $\rho'$? We cast this question as an SDP feasibility problem and we employed a variant of Farkas' Lemma to find the necessary and sufficient conditions for the existence of such an operation. Then, we provided a set of magic monotones which we proved to be complete. In other words, the answer to the single-shot question is positive in our case if and only if the density matrix $\rho\in\Den$ is more resourceful than (or at least as resourceful as) the density matrix $\rho'\in\Denp$ according to all the magic monotones~\eqref{def-monotone}. Note that we answered this question for the exact conversion of a resource state to another resource state. As a future line of research, the possibility of approximate conversion of a magic state to another using SPOs is a natural next step.\\

The usefulness of the tools developed here may extend beyond the study of magic states. As a future line of research, it would be interesting to see if the methods applied in this paper (such as SDP and Farkas' Lemma) to the resource theory of magic can be applied to the more general case of convex resource theories.\\

Another possible direction for future work is to extend our results to the case of infinite-dimensional systems, namely the resource theory of non-Gaussianity. It is known that stabilizer states are the discrete analogues of Gaussian states~\cite{Gross} and  that non-Gaussianity is a resource for tasks such as distillation of Gaussian entanglement, violation of continuous-variable Bell inequalities  and continuous-variable quantum computation~\cite{Genoni2010}. However, it is still unknown what can be achieved using Gaussian operations given that one has access to a non-Gaussian state.    

\acknowledgments
We thank Mercedes Gimeno-Segovia, Mark Girard and David Jennings for useful discussions and comments. The authors were supported by the Natural Sciences and Engineering Research Council of Canada (NSERC) and the Pacific Institute for the Mathematical Sciences (PIMS). BCS appreciates financial support from Alberta Innovates Technology Futures and China's 1000 Talent Plan.

\appendix
\section{Numerical examples}
Here, we present an example for the qutrit case where the state $\rho$ cannot be converted using the stabilizer operations as was defined in \cite{Veitch2014}, while this conversion is possible using SPO. To find such examples, we generated random density matrices $\rho$ and $\rho'$ and ordered them such that $\rho'$ has a higher sum negativity in comparison to $\rho$. This means that the state $\rho$ cannot be converted to $\rho'$ using the stabilizer operations defined in \cite{Veitch2014}. Then, using the CVX package for Matlab \cite{cvx} (a package for specifying and solving convex programs), we checked whether there exists an SPO with the Choi matrix $J$ which satisfies conditions \eqref{SDP}.\footnote{https://github.com/ahmadimehdi/Quantification-and-manipulation-of-magic-states.}\\

As an example for the qutrit case, consider the following density matrices:
\begin{widetext}
\begin{equation}
\rho =
  \left[ {\begin{array}{ccc}
   \phm 0.1913 + 0.0000i &  \phm 0.0580 - 0.1002i  & \phm 0.1383 + 0.1163i \\
   \phm 0.0580 + 0.1002i &  \phm 0.4585 + 0.0000i  & -0.0292 - 0.0897i \\
   \phm 0.1383 - 0.1163i  & -0.0292 + 0.0897i & \phm 0.3502 + 0.0000i 
  \end{array} } \right]\nonumber
\end{equation} 
and 
\begin{equation}
\rho'=
  \left[ {\begin{array}{ccc}
   \phm 0.2383 + 0.0000i  & \phm 0.0413 + 0.0808i & -0.0286 - 0.0380i \\
   \phm 0.0413 - 0.0808i  & \phm 0.4894 + 0.0000i  & \phm 0.1650 - 0.0552i \\
  -0.0286 + 0.0380i  & \phm 0.1650 + 0.0552i  & \phm  0.2723 + 0.0000i \\
  \end{array} } \right].\nonumber
\end{equation} 
The sum-negativities of these two states are $0$ and $0.0074$ respectively, while there exists an SPO that converts $\rho$ to $\rho'$, whose Choi matrix has the real part
\begin{equation}
J_{\text{re}}=
  \left[ \begin{array}{ccccccccc}
\phm 0.3841  & -0.0380  & -0.0629  & -0.0173  & -0.0242  & \phm 0.0782  & \phm 0.0270  & -0.0073  & -0.0597  \\ 
-0.0380  & \phm 0.2577  & \phm 0.0535  & \phm 0.0516  & \phm 0.0313  & -0.0470  & -0.0406  & -0.0232  & \phm 0.0114  \\ 
-0.0629  & \phm 0.0535  & \phm 0.2688  & -0.0287  & \phm 0.0285  & \phm 0.0243  & -0.0013  & -0.0241  & -0.0081  \\ 
-0.0173  & \phm 0.0516  & -0.0287  & \phm 0.2935  & \phm 0.0442  & \phm 0.0853  & -0.0711  & \phm 0.0901  & \phm 0.0663  \\ 
-0.0242  & \phm 0.0313  & \phm 0.0285  & \phm 0.0442  & \phm 0.4680  & -0.0595  & \phm 0.0460  & \phm 0.1288  & -0.0375  \\ 
\phm 0.0782  & -0.0470  & \phm 0.0243  & \phm 0.0853  & -0.0595  & \phm 0.4349  & \phm 0.1283  & -0.0815  & \phm 0.1109  \\ 
\phm 0.0270  & -0.0406  & -0.0013  & -0.0711  & \phm 0.0460  & \phm 0.1283  & \phm 0.3224  & -0.0062  & -0.0224  \\ 
-0.0073  & -0.0232  & -0.0241  & \phm 0.0901  & \phm 0.1288  & -0.0815  & -0.0062  & \phm 0.2742  & \phm 0.0060  \\ 
-0.0597  & \phm 0.0114  & -0.0081  & \phm 0.0663  & -0.0375  & \phm 0.1109  & -0.0224  & \phm 0.0060  & \phm 0.2964  \\ 

 \end{array} \right]  \nonumber
\end{equation}
and imaginary part
\begin{equation}
J_{\text{im}}=
  \left[ \begin{array}{ccccccccc}
\phm0.0000  & -0.0318  & \phm 0.0458  & -0.0795  & \phm 0.0678  & \phm 0.0427  & \phm 0.0641  & -0.0562  & -0.0273  \\ 
\phm 0.0318  & \phm0.0000  & -0.0306  & \phm 0.0295  & \phm 0.0548  & -0.0241  & -0.0270  & -0.0188  & \phm 0.0147  \\ 
-0.0458  & \phm 0.0306  & \phm0.0000  & \phm 0.0902  & -0.0621  & \phm 0.0500  & -0.0629  & \phm 0.0406  & -0.0269  \\ 
\phm 0.0795  & -0.0295  & -0.0902  & \phm0.0000  & \phm 0.0508  & -0.0664  & \phm 0.0425  & \phm 0.0302  & -0.1252  \\ 
-0.0678  & -0.0548  & \phm 0.0621  & -0.0508  &\phm 0.0000  & \phm 0.0475  & -0.0945  & -0.0393  & \phm 0.0826  \\ 
-0.0427  & \phm 0.0241  & -0.0500  & \phm 0.0664  & -0.0475  & \phm0.0000  & \phm 0.0398  & -0.0373  & -0.0368  \\ 
-0.0641  & \phm 0.0270  & \phm 0.0629  & -0.0425  & \phm 0.0945  & -0.0398  & \phm0.0000  & -0.0190  & \phm 0.0205  \\ 
\phm 0.0562  & \phm 0.0188  & -0.0406  & -0.0302  & \phm 0.0393  & \phm 0.0373  & \phm 0.0190  & \phm0.0000  & -0.0169  \\ 
\phm 0.0273  & -0.0147  & \phm 0.0269  & \phm 0.1252  & -0.0826  & \phm 0.0368  & -0.0205  & \phm 0.0169  & \phm0.0000  \\ 

 \end{array} \right]  \nonumber
\end{equation}

corresponds to an SPO that converts $\rho$ to $\rho'$.
\end{widetext}
\bibliography{ResourcetheroyofNSS}

\end{document}